# Numerical model of Electron Cyclotron Resonance Ion Source

V. Mironov, S. Bogomolov, A. Bondarchenko, A. Efremov, V. Loginov

*Joint Institute for Nuclear Research, Flerov Laboratory of Nuclear Reactions, Dubna, Moscow Reg. 141980, Russia*

Important features of Electron Cyclotron Resonance Ion Source (ECRIS) operation are accurately reproduced with a numerical code. The code uses the particle-in-cell technique to model a dynamics of ions in ECRIS plasma. It is shown that gas dynamical ion confinement mechanism is sufficient to provide the ion production rates in ECRIS close to the experimentally observed values. Extracted ion currents are calculated and compared to the experiment for few sources. Changes in the extracted ion currents are obtained with varying the gas flow into the source chamber and the microwave power. Empirical scaling laws for ECRIS design are studied and the underlying physical effects are discussed

## I. Introduction

Electron Cyclotron Resonance Ion Source (ECRIS) is a plasma-based device designed to produce intense beams of multiply charged ions [1]. Plasma in the source is a microwave low-pressure ($10^{-8}$-$10^{-6}$ mbar) discharge in a static magnetic field of a few Tesla. Typically, microwave power is in the kW range and frequency of the microwaves is of a few GHz, with the modern ECRIS designs aimed to 56 GHz or higher. Electrons in the plasma are heated by absorption of the microwaves at the electron cyclotron resonance surface, where the electron cyclotron frequency equals to the microwave frequency. The resonance magnetic field of 0.5 T corresponds to the 14 GHz microwave frequency. The ECR surface should be closed and not touching the vacuum chamber walls to get the plasma electron temperature around 1 keV favorable for producing the highly charged ions. Electrons in the ECRIS plasma are confined by the mirror magnetic forces. The so-called minimum-B magnetic structure is needed that is formed by a set of solenoids to produce a longitudinal magnetic trap combined with a multipole magnetic field increasing toward the walls in the radial direction. To form the multipole field, either a set of permanent magnets in the hexapole Halbach configuration or a set of the superconducting coils is used. The minimum-B structure provides a favorable curvature of the magnetic field lines inside the plasma capable to suppress the plasma macro-instabilities.

In Fig.1, the magnetic field lines are visualized for a typical ECRIS. Only those lines are shown that cross the ECR zone, where the main plasma production takes the place. The lines are terminated at the walls of a cylindrical vacuum chamber. The ECR zone is colored white. Since the plasma flows predominantly along the magnetic field lines, this picture also depicts the plasma shape in ECRIS with its 120-degree symmetry.

Along the axis of plasma symmetry (z-axis), plasma is limited by two flanges at the "injection" and "extraction" sides of the source. At the injection flange, microwaves are coupled into the plasma chamber through a waveguide with an open end. Also, a gas inlet is placed there, as well as the axial negatively biased electrode used to control a plasma potential spatial distribution [2]. At the extraction flange, ions leave the source through an extraction aperture. Normally, sources are positively biased in respect to a ground with up to a few tens of kV voltage for the ion beam formation and acceleration. Extracted ion beams are then used in variety of applications, e.g. for injection into particle accelerators. It is worth to note that this is an ECRIS source that is used as a particle injector for the heavy ion program at the LHC.

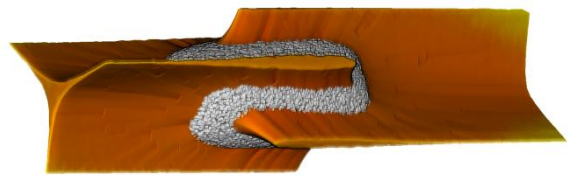

Fig.1. Bird's-eye view of the magnetic field lines crossing the ECR zone in ECRIS. The ECR zone is colored white.

Theoretical models for the ECRIS operation began to be developed soon after the source invention in the eighties of last century. The dimensionless models solved a set of nonlinear equations for a balance of ion and electron densities inside the plasma taking into account particle production and loss processes [3].

Ion lifetime (confinement time) plays an important role in such the equations defining the mean charge state of ions achievable for a given electron density. It has been commonly argued that a small potential dip should be formed inside the ECR zone in respect to the globally positive plasma potential to balance the ion and electron losses out the ECRIS plasma [4]. Ion production volume is limited predominantly by the ECR zone, and due to the fast ion-ion collisions, ions with the different charge states are thermally equilibrated. Potential dip value is assumed to be large compared to the ion temperature. Ion confinement times in these conditions have the exponential dependence on the ion charge state Q.

At least two experimental results are in contradiction with the potential dip concept. Douysett et al. [5] measured the spectrally resolved X-ray emission from argon ECRIS plasma. From the line intensities, ion densities were obtained and compared with the extracted ion currents. Weak linear dependence of the confinement times on the ion charge state was observed. Pulsed injection of metal atoms into ECRIS has been studied by using the laser ablation technique [6]. Time structure of the extracted ion currents was not consistent with the exponential dependence of the ion confinement times on Q.

It was shown elsewhere [7] that reasonable agreement with the experimentally observed features of ECRIS can be obtained without involving the potential-dip ion confinement mechanism. The present work reports on further development of the model, solving some problems in interpretation of the ECRIS behavior and reaching the predictive level. The model is based on the particle-in-cell algorithm with the Monte-Carlo collision block to model a dynamics of ions inside the plasma. Electrons in the model are treated as a neutralizing background, with the electron temperature taken as an input.

This paper is organized as follows. In Section II parameters of the modeled ECR sources are given and the model is described. Section III reports the results of calculations concerning a general source behavior, such as the source responses to variations in the microwave power and gas flow into the source chamber. Spatial distributions of the plasma inside the source and of the ion fluxes to the walls are discussed, ion confinement times are estimated. Section IV presents the source output variations with changing the magnetic field parameters, such as the hexapole ($B_H$) and the longitudinal magnetic field extreme values – magnetic fields at injection ($B_{Inj}$), extraction ($B_{Ext}$) and minimum magnetic field ($B_{Min}$). Conclusions are given in the end.

## II. The model description

The code uses the standard particle-in-cell technique to trace a movement of a large number of macro-particles representing heavy ions and atoms. Calculations are done on the Cartesian computational mesh of 65x65x64 cells in x,y and z directions respectively. Total number of macro-particles is $2 \times 10^5$ with a statistical weight of a macro-particle being an input for the code. The weight varies in the range from $10^7$ to $10^{10}$ real particles per macro-particle. In typical conditions, neutral particles constitute around 75% of the total particle number. Atoms move straight in the line inside the cylindrical plasma chamber and are reflected back when hitting the chamber walls. Angles of reflection are selected according to the cosine-law for the perfectly diffusing walls. Atom velocities are selected from the Maxwell-Boltzmann velocity distribution with the gas temperature of 0.025 eV (room temperature); velocity is selected whenever an atom hits the wall or when atom is injected into the plasma chamber. Particles leave the system through the round aperture at the extraction side of the source. After leaving the chamber, particles are returned back into the computational domain with the initial coordinates that correspond to the gas inlet position. Thus, the total number of particles in the domain is always kept constant. The result is that only the stationary processes are calculated properly, while the transient processes need to be treated in a different manner [8].

The calculations deal exclusively with argon as a working gas. Inclusion of other elements is possible but is not considered as needed at the moment.

Particles are moving in the magnetic field of the source. The code uses the source dimensions (plasma chamber length and diameter) and magnetic field distributions (including the length of the ECR zone $L_{ECR}$) of four ECR ion sources – KVI AECRIS from KVI, Groningen [9], ECR-4M[2] [10], DECRIS-2M [11] and DECRIS-SC2 [12] sources from Flerov Laboratory for Nuclear Reactions (FLNR), JINR. Parameters of the sources are listed in Table I. DECRIS-SC2 source is modeled for two microwave frequencies of 14 and 18 GHz, which implies different tunings of the magnetic field while keeping the source geometry the same. Other sources use 14 GHz microwaves. The magnetic field values are shown for the tunings of the sources optimized to produce maximal $Ar^{8+}$ ion beams. The hexapole fields had been measured at the radii that correspond to the inner size of plasma chamber.

Table I. Plasma chamber dimensions and magnetic field extremes of the modeled sources.

|  | KVI-AECRIS | ECR4-M$^2$ | DECRIS-2M | DECRIS-SC2 14 GHz | DECRIS-SC2 18 GHz |
|---|---|---|---|---|---|
| Length, cm | 30 | 18 | 20 | 28 | 28 |
| Diameter, cm | 7.6 | 7.4 | 6.4 | 7.4 | 7.4 |
| $B_{Inj}$, T | 1.97 | 1.1 | 1.21 | 1.66 | 1.97 |
| $B_{Ext}$, T | 1.07 | 1.09 | 1.01 | 1.08 | 1.35 |
| $B_{Min}$, T | 0.35 | 0.38 | 0.4 | 0.35 | 0.47 |
| $B_H$, T | 0.85 | 1.0 | 1.0 | 1.1 | 1.1 |
| Extraction aperture, cm | 0.8 | 1.0 | 1.0 | 1.0 | 1.0 |
| $L_{ECR}$, cm | 10.8 | 6.7 | 6.2 | 7.3 | 7.4 |

Though the sources have different sizes of the plasma chamber and magnetic field distributions, their outputs are comparable. Current of the extracted $Ar^{8+}$ ions varies from 0.4 to 0.7 mA, being maximal for the DECRIS-SC2 18 GHz. The sources can be considered as well-performing typical representatives of the ECRIS 2$^{nd}$ generation. All sources use the permanent magnets to form the hexapole field. Four superconducting coils form the longitudinal trap in DECRIS-SC2, other sources have two solenoids. Vacuum chamber of KVI-AECRIS is made of aluminum, the FLNR source chambers are made from stainless steel.

To calculate the longitudinal magnetic field components ($B_r$, $B_z$) we use the Poisson-Superfish code [13] for an axially symmetric set of solenoids in combination with soft iron plugs and yokes. Typical geometry of the magnetic system is shown in Fig.2 for the KVI-AECRIS source.

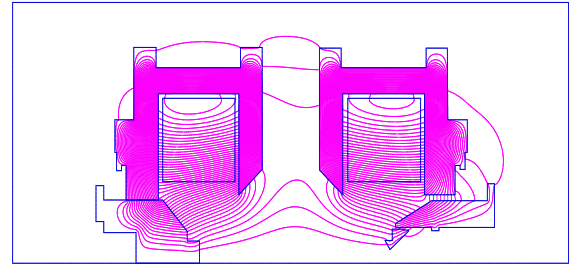

Fig.2. Geometry of the longitudinal magnetic trap for the KVI-AECRIS source.

At the figure, extraction side of the source is at the left. We distinguish between the extraction (left) and the injection (right) solenoids. In this geometry, the magnetic field along z-axis behaves as it is shown in Fig.3. The extremes of the field $B_{Inj}$, $B_{Ext}$ and $B_{Min}$ are labeled at the graph, as well as the resonant magnetic field $B_{ECR}$=0.5 T.

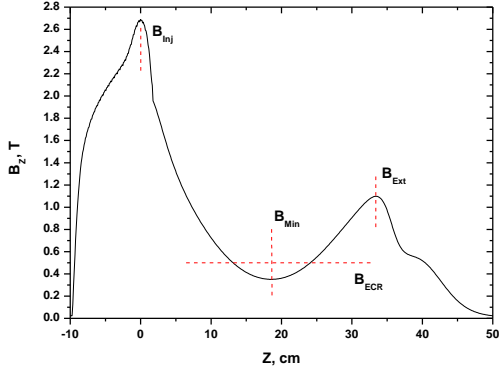

Fig.3. Magnetic field at the axis of KVI AECRIS.

Magnetic field of the source is a sum of the longitudinal field and the hexapolar component. In our model, the hexapolar field is calculated analytically neglecting the edge effects. Then the radial $B_x$ and $B_y$ components of the total magnetic field are calculated as follows:

$$B_x = B_r \cos(\theta) + \frac{B_H \sin(2\theta) R^2}{R_0^2}$$

$$B_y = B_r \sin(\theta) + \frac{B_H \cos(2\theta) R^2}{R_0^2}$$

where θ is the polar angle, R is radius, $B_H$ is the hexapolar field at the plasma chamber wall and $R_0$ is the chamber radius. The longitudinal component $B_z$ is fully determined by the solenoidal magnetic field.

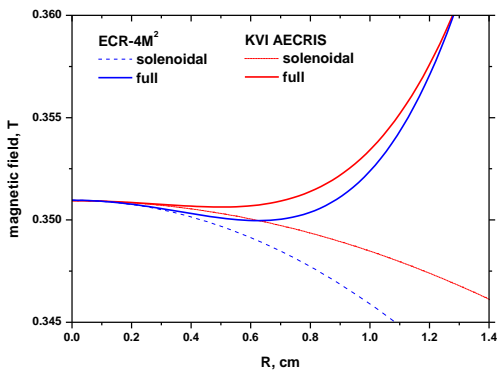

Fig.4. Radial dependence of the magnetic field at the plane where the magnetic field at the source axis reaches its minimum. Fields of ECR-4M$^2$ are shown as the blue lines; red lines are for the KVI AECRIS

At the plane where the solenoidal magnetic field at the axis of symmetry (z-axis) reaches its minimum, the radial component $B_r$ of the solenoidal field is zero and $B_z$ component of the solenoidal field decreases with radius. The total magnetic field (solenoidal + hexapolar) at this plane is displayed at Fig.4 for two sources simulated by the code, KVI AECRIS and ECR-4M$^2$. The $B_z$ solenoidal components are shown by the dashed lines.

Fields are shown close to the axis. For KVI AECRIS the radial gradient of the solenoidal field is smaller than the field of ECR-4M$^2$ and other FLNR sources. The result is that full (solenoidal + hexapole) field is smoothly increasing with radius for KVI AECRIS but have a local minimum at R≈0.7 cm for ECR-4M$^2$, DECRIS-2M and DECRIS-SC2 sources. This influences the plasma shape in the way that will be discussed later.

When moving in the magnetic field, ions experience elastic and inelastic collisions with each other and with electrons. The Monte-Carlo collision block of the code includes the following processes: ion-ion collisions, ion heating and diffusion due to the electron-ion collisions, elastic and inelastic collisions of ions with atoms, ionization and recombination.

The fastest process in the collision block is the ion scattering in ion-ion collisions. Frequency of the collisions is comparable or exceeds the Larmor frequency of ion rotation in the magnetic field. Collision frequency is calculated as given in [14]. The collisions are treated by using the standard Takizuke-Abe method of pairing the collision partners in a cell [15]. The method ensures the energy and momentum conservation for ions. Angle of scattering is calculated each time step according to the Nanbu algorithm [16]. The Coulomb logarithm $\lambda_{\alpha\beta}$ for the ion-ion collisions is chosen to be constant and equal to 10, close to values given by Huba [14] for the mixed ion-ion collisions.

Cross-sections for the elastic and inelastic collisions of singly charged argon ions with argon atoms are taken from [17,18]. For the multiply charged ions, we scale the cross-sections of the singly charged ions linearly with the ion charge state Q. The elastic scattering in ion-atom collisions is treated as isotropic. For the charge transfer, we consider only the single-electron transfer neglecting the multiple-electron transfer processes. The charge transfer for the doubly charged argon ions is energetically forbidden. After charge-transfer for the multiply charged ions, kinetic energy release results in an energization of the colliding particles. The energy release with the typical value of 10 eV is considered to be equally distributed between the collision partners. Ion-atom collisions are included for the sake of completeness; they do not influence the ion dynamics in ECRIS substantially, mainly because of a very low atom density inside the plasma.

Ion heating and scattering in ion-electron collisions are computed by kicking an ion each time step in (almost) random direction with the velocity diffusion coefficient D.

$$V_i = V_{i0} + D(R+\delta)\frac{|R|}{|R+\delta|}$$

Here, $R$ is a random vector and the velocity diffusion coefficient is calculated as

$$D = \left(\frac{3.715\times10^{-6} n_e \lambda_{ei} Q dt}{M_i^2 \sqrt{T_e}}\right)^{1/2}$$

Such the procedure is equivalent to the classical ion heating due to the electron-ion collisions with the heating rate from Huba. The heating rate is directly proportional to the electron density at the given computational cell, quadratically depends on the ion charge state and is inversely proportional to the square root of the electron temperature – the colder are electrons, the more efficiently they heat the ions, providing that the electron temperature remains much larger than the ion temperature. The Coulomb logarithm $\lambda_{ei}$ for the ion-electron collisions is set to 17.

Direction of the kick in ion velocity is not random in the calculations. For the fully ionized plasma in a strong magnetic field, the resistively driven flow anti-parallel to the pressure gradient should be taken into account, $u_\perp = -\frac{\eta}{B^2}\nabla_\perp p$, where $u_\perp$ is the drift velocity, $\nabla_\perp p$ is the pressure gradient perpendicular to magnetic field, B is magnetic field and $\eta$ is the plasma resistivity [19].

Electrons in ECRIS plasmas are strongly magnetized and rotating around the magnetic field lines. Ions are preferentially pushed by the electron-ion collisions in the direction of electron diamagnetic drift, which is perpendicular to the direction of the electron (ion) density gradient. In the strong magnetic field, the ions are then driven in the $F\times B$ direction, where $F$ is the force acting on the ions and $B$ is the magnetic field. To account for this effect, we estimate the electron density radial gradient on the computational mesh, scale it with the electron Larmor radius and generate a vector $\delta$ in the direction of the electron diamagnetic drift for the given mesh cell. Only x and y components of $\delta$ are calculated, z-component is small compared to other components.

$$\delta_x = -\frac{R_L}{n_e(ix,iy,iz)}\begin{pmatrix} -n_e(ix,iy+1,iz)+n_e(ix,iy-1,iz)+ \\ \frac{1}{2}\begin{pmatrix} -n_e(ix-1,iy+1,iz)-n_e(ix+1,iy+1,iz)+ \\ n_e(ix+1,iy-1,iz)+n_e(ix-1,iy-1,iz) \end{pmatrix} \end{pmatrix}$$

$$\delta_y = -\frac{R_L}{n_e(ix,iy,iz)}\begin{pmatrix} -n_e(ix-1,iy,iz)+n_e(ix+1,iy,iz)+ \\ \frac{1}{2}\begin{pmatrix} -n_e(ix-1,iy+1,iz)+n_e(ix+1,iy+1,iz)+ \\ n_e(ix+1,iy-1,iz)-n_e(ix-1,iy-1,iz) \end{pmatrix} \end{pmatrix}$$

$$\delta_z = 0$$

Here, scaling coefficient $R_L$ is the electron Larmor radius scaled with the mesh step size.

$$R_L = \frac{2.38\times10^{-6}\sqrt{T_e}}{B(ix,iy,iz)\Delta X}$$

For the typical plasma conditions, amplitude of $\delta$ is much smaller than 1. In the next step, we generate a random vector $R$ and sum it with $\delta$. The calculated vector after renormalization defines the direction and amplitude of the ion kick due to the electron-ion collisions.

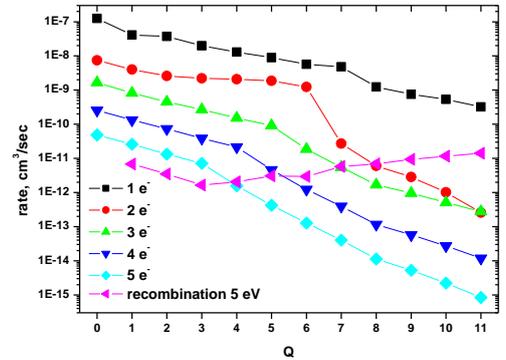

Fig.5 Rates for single and multiple ionization of argon atoms and ions for the electron temperature of 1 keV. Recombination rate is shown for the electron temperature of 5 eV.

Probability for a particle to change its charge state is calculated each time step depending on the electron density and temperature in the given computational cell. The single ionization rates are combined from datasets [20,21]. Double ionization rates are taken from the fit of [22] after correcting for some mistypes in their table of fitting coefficients. For the ionization rates with a number of removed electrons n>=3 we use the scaling from [23]. Recombination processes are included for completeness, though they are too small [21] for the considered plasma parameters and ion charge states to influence the plasma dynamics noticeably. In Fig.5, the ionization rates are shown for the electron temperature of 1 keV for argon ions with the charge state from 0 to 11. Double ionization is important; the processes with n≥3 are minor contributors to the full ionization rates. Error bars for the rates are not shown in the graph, but they can be large for some charge states.

Just before hitting the walls of the plasma chamber, the charged particles are accelerated in a plasma sheath up to the substantially high energies around (20-50)×Q eV. We assume full neutralization of ions in collisions with the walls and full accommodation of their energy. Neutralized particles are reflected back into the chamber diffusively with the velocities selected from the Maxwell-Boltzmann distribution with the room temperature. Generally speaking, depending on the atomic masses of particles and wall material, wall conditions, angle of incidence and other factors, the reflected particles retain some their initial energy and momentum. In our simulations, however, we take the thermal accommodation factor equal to 1 and use the "cosine-law" for calculation of the scattering angles. According to the experimental observations, this is justified for the cases when gas atomic mass is higher than the mass of wall atoms [24]. Argon accommodation coefficient for the aluminum surface is reported [25] to be 0.8-0.9.

Those particles that leave plasma chamber through the round aperture at the extraction side of source form the charge state distribution of the extracted ion currents. The gas flow through the extraction aperture should be equal to the flow into the system. Gas pressure distribution inside the plasma chamber varies so much that it cannot be used as a global parameter that describes the source conditions. Instead, we use the gas flow out of the chamber to parameterize the source operation. Most of the particles flowing out of the chamber are ions; atom flux is less than 15% of total value in the typical source conditions.

In calculations the electron density is always assumed to be equal to the ion charge density from requirement of charge neutrality. Electron energy distribution function (EEDF) is supposed to be Maxwellian with the electron temperature taken as an input for the code. This is a simplification, since it is experimentally known that the EEDF deviates from the Maxwell-Boltzmann distribution significantly. X-ray measurements [26] indicate that there is a high-energy tail in the distribution with the "spectral temperature" of 25-50 keV. In the following, we neglect these "hot" electrons, which anyway do not contribute into the ionization and heating of the ions in the plasma due to their low collisionality.

We divide the plasma electrons into two spatially separated components: the "warm" electrons inside the ECR zone are assumed to have a temperature $T_{ew}$ in keV range, the electrons outside the zone are assumed to be "cold" with the fixed temperature $T_{ec}$ of 5 eV. The cold electron temperature can vary in a wide range from 1 to 10 eV without affecting the code outputs significantly. When cold electron temperature exceeds 10 eV in calculations, $Ar^{1+}$ ion current is much higher than the $Ar^{2+}$ current, which is not observed in the experiments. In un-magnetized argon plasma with electron temperature of 5 eV, plasma positive potential should be roughly 25 V to balance the ion and electron losses to the plasma conducting boundaries. Such the plasma potential values are typical for the ECRIS plasmas. Electrostatic probe measurements of the cold electron temperature in peripheral ECRIS plasma give the same estimates of a few eV. Localization of the warm electrons inside the ECR zone follows from the dynamics of electron heating by microwaves perpendicular to the magnetic field lines and electron trapping by the magnetic mirror forces. Such the localization was confirmed by the numerical simulations reported elsewhere [27].

When performing the simulations of ECRIS, the total losses of ions to the chamber walls and into the extraction aperture are calculated. These losses should be equilibrated with exactly the same flux of electrons out of the plasma. The typical current of electrons to the walls is around 30 mA in our conditions. The power deposition to the walls due to electron losses is $P_e = \frac{3}{2}kT_{ew}$; summing this value with the potential and kinetic energies of the lost ions gives us the total power losses out of the plasma. Other power loss channels, such as X-ray and electron cyclotron emission are negligibly small in our conditions. From the power balance, it follows that the power losses from the plasma are equal to the microwave power coupled to the plasma. These estimates give the lower limit for the coupled power, since we do not account for the power losses connected with the high-energy tail of the electron energy distribution. Connection of the injected microwave power to the coupled power is also not straightforward, since there are losses of microwaves at the chamber walls and the microwave reflection.

For the given configuration of ECRIS magnetic field, the model requires only two input parameters: statistical weight of the macro-particles and the warm electron temperature. When solution is converged, the applied inputs result in the plasma that can be characterized with two output parameters: gas flow out/into the plasma chamber and the power losses out of the plasma (coupled microwave power). Normally we perform the calculations by adjusting the inputs such as to reach the required output parameters (±10%), e.g. by fixing the power losses at some level and then varying the gas flow in some range to study the source response to the gas flow variations.

In the following sections, we describe the results of applying the model for simulations of ECRIS, starting with studying the general features of the sources.

## III. General characteristics.

### A. Charge state distributions of the extracted ions.

The charge state distribution (CSD) of the extracted ion currents is one of the most important characteristics of any ECRIS. It quantifies the ion beam intensity extracted from the source for the given ion charge states. The simulated CSD for KVI-AECRIS is shown in Fig.6 for the source conditions optimized to produce the maximal current of $Ar^{8+}$ ions. At this, the warm electron temperature is 2 keV and the particle statistical weight was chosen to reach the total power coupled to the plasma equal to 100 W.

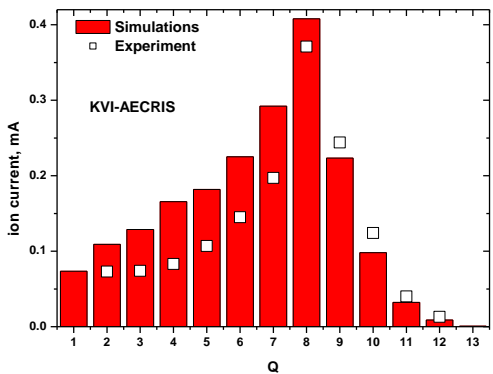

Fig. 6. Charge state distribution of the extracted ion currents for KVI-AECRIS. $T_{ew}$=2 keV, P=100 W. Experimental points are shown as open squares.

The experimental data at Fig.6 were obtained at the injected microwave power of 800 W and at the gas flow into the plasma chamber tuned to maximize $Ar^{8+}$ current. An overall satisfactory agreement is seen between the simulations and the experiment, with some over-estimation of the calculated low charge state currents. Note here that the experimental points are not corrected for the beam losses in the low-energy beam line, which can be as high as 30% [28]. These losses as well as the uncertainties in the ionizations rates may account for the observed deviations.

For all other simulated sources, the same level in reproduction of the experimental charge state distributions is achieved. For three FLNR sources, the best correspondence is obtained with setting the coupled microwave power to 200 W, compared to 100 W for the KVI-AECRIS source. Optimal electron temperature for these sources is slightly higher compared to KVI-AECRIS, 3 keV instead of 2 keV.

### B. Shapes of the plasma and of the extracted ion beams.

The basic assumption of the model about localization of the warm electron component in the ECR zone defines the plasma shape. Electron density is peaked on the source axis and has a maximum close to the B-min position along z-axis. This is illustrated by Fig.7 and 8, where the electron density is shown as a function of transversal coordinate x at the center of the plasma (z=17 cm), and along z-axis at x=y=0 cm.

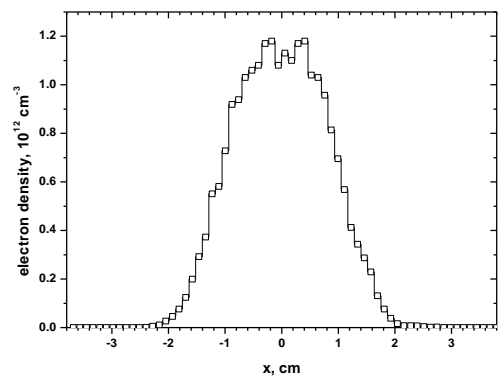

Fig.7. Electron density as a function of radius for KVI AECRIS (z=17 cm).

Direction of z-coordinate is chosen such that the injection side of the source is at z=0. The calculations are done for KVI-AECRIS tuned for the $Ar^{8+}$ production. The source parameters are chosen the same as presented in Fig.6.

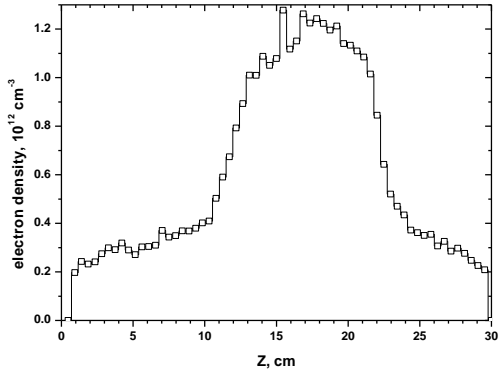

Fig.8. Electron density as a function of longitudinal coordinate (x=y=0 cm). KVI AECRIS.

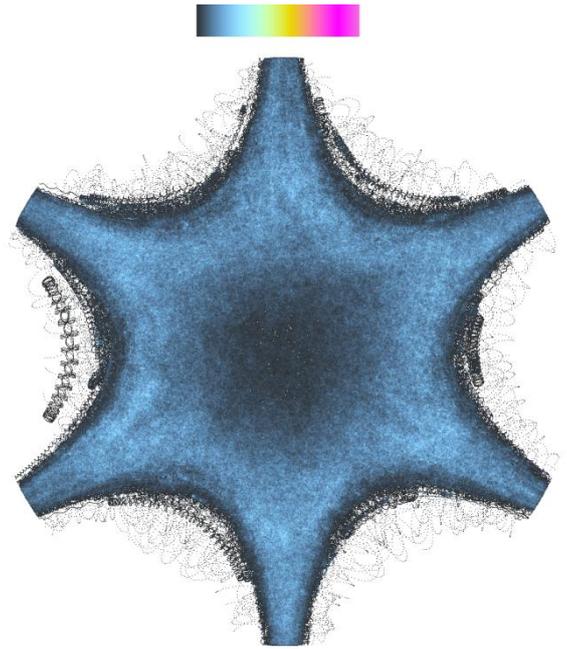

Fig.10. Spatial distribution of argon ions $Ar^{1+}$ in the transversal x-y plane close to the maximum in the plasma density along the source longitudinal axis. KVI AECRIS.

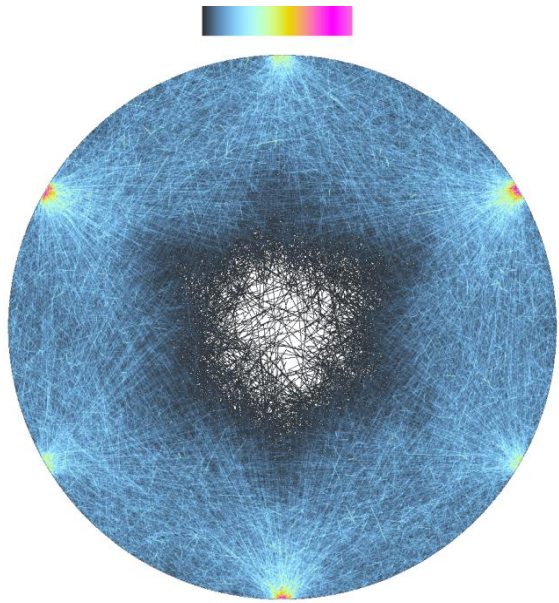

Fig.9. Spatial distribution of argon atoms $Ar^0$ in the transversal x-y plane close to the maximum in the plasma density along the source longitudinal axis. KVI AECRIS

Outside the ECR zone, the electron density decreases fast. The maximal density of the distribution is around $1\times10^{12}$ cm$^{-3}$, well below the critical density of $2.5\times10^{12}$ cm$^{-3}$ for the 14 GHz microwave frequency.

In Figures 9-11 we show the x-y spatial distribution of ion and atom density at z=17±0.5 cm close to position of $B_{min}$ along the source axis. The pictures are obtained by tracing the particle movement and incrementing a screen pixel color index by 1 whenever a particle hits the pixel. Color scale is shown in an upper part of the pictures and corresponds to the number of the hits from 1 to 250. Times of exposition are different for the shown distributions.

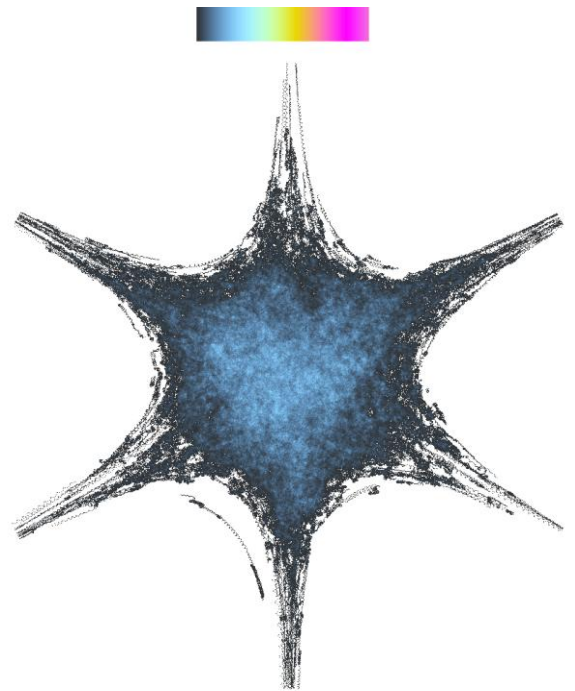

Fig.11. Spatial distribution of argon ions $Ar^{8+}$ in the transversal x-y plane close to the maximum in the plasma density along the source longitudinal axis. KVI AECRIS.

The six-arm symmetry is seen in the Figures. The gas density decreases fast in the direction toward the chamber center, where the plasma density is high. The highest atom density is close to the chamber walls where ions are reflected back into the chamber after neutralization. These positions are visible as the bright red spots in Fig.9. The "hole" in the distribution is due to effective ionization of neutral argon by electron impact in the plasma. The ionization rate for neutral argon atoms is around $1\times10^{-7}$ cm$^3$/sec, the atom thermal velocity for the $T_0$ =0.025 eV is $4\times10^4$ cm/sec. For the electron density of $5\times10^{11}$ cm$^{-3}$ the mean free path of argon atoms before ionization is ≈1 cm - new ions are produced from neutral atoms almost exclusively at the ECR zone periphery. Transport of ions into the central parts of plasma across the magnetic field lines is slow. In these conditions, if gas mean velocity is increased in some way, this results in a substantial increase of the extracted ion currents. Also decreasing the ECR zone size facilitates a penetration of atoms into the plasma dense regions.

Spatial distributions of the lowly charged ion are hollow in the radial direction (Fig.10). The higher is the charge state of ions, the stronger is localization of ions close to the source axis.

In more quantitative way, the spatial distributions of particle densities are shown in Fig.12 for the radial dependence and Fig.13 for the dependence on z-coordinate along the plasma axis of symmetry (x=y=0).

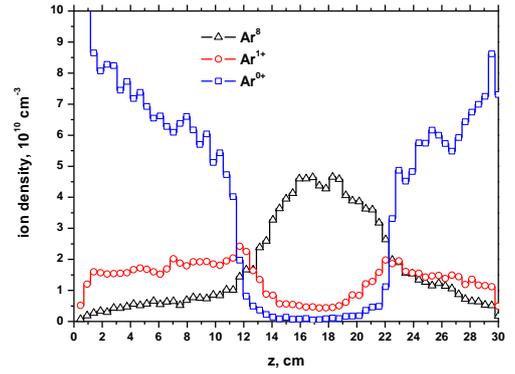

Fig.13. Ion and atom density as a function of z coordinate along axis of symmetry of the plasma. $T_{ew}$=2 keV. KVI AECRIS.

It is seen from Figs.12-13 that the atom density drops by more than order of magnitude when comparing the regions outside the dense plasma to the plasma center. The ion densities decrease fast outside the ECR zone both in radial and axial directions, the highest charge states are peaked at the source axis. The distributions are obtained with the warm electron temperature $T_{ew}$=2 keV and the coupled microwave power of 100 W.

The distributions vary with changing the gas flow into the source chamber and electron temperature. In Fig.14, the same distributions are shown as in Fig.12, but for the lower warm electron temperature $T_{ew}$=900 eV. The coupled microwave power is still 100 W.

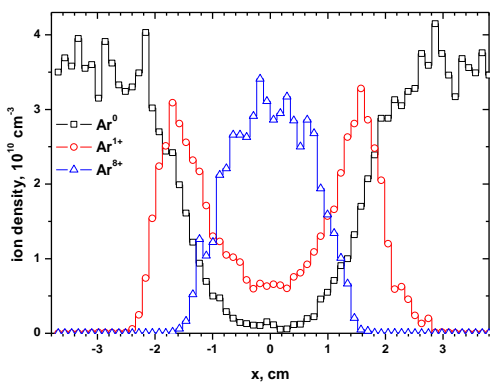

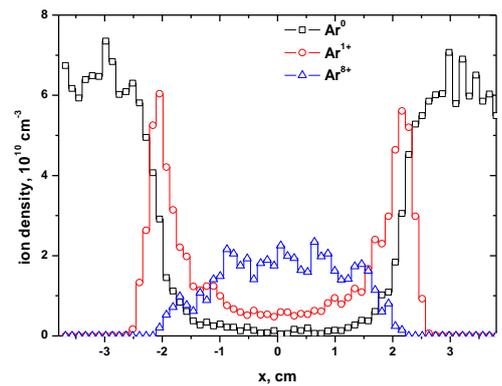

Fig.12. Ion and atom density as a function of x at the transversal plane close to the $B_{min}$ position in z-direction (z=17 cm). $T_{ew}$=2 keV. KVI AECRIS.

Fig.14. Ion and atom density as a function of x at the transversal plane close to the B-min position in z-direction (z=17 cm). $T_{ew}$=900 eV. KVI AECRIS.

The particle density in Fig.12 is shown along x-axis; this is a density profile in the horizontal direction of the distributions shown in Figs.9-11.

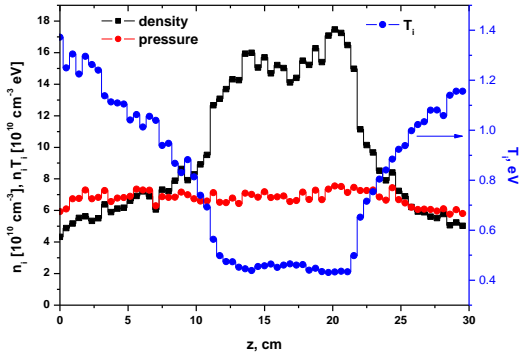

Fig.15. Total ion density and ion pressure as a function of z coordinate along axis of symmetry of the source. $T_{ew}$=2 keV. Ion mean temperature is also shown (blue circles, right scale). KVI AECRIS.

Comparing Figs.12 and 14, we see that the gas density outside the ECR zone is substantially higher for the source conditions with the lower warm electron temperature (approximately $6\times10^{10}$ cm$^{-3}$ vs. $3.5\times10^{10}$ cm$^{-3}$), while the highly charged ion density drops at the source axis by a factor of almost 50%. Changes in the extracted ion currents due to such broadening of the plasma will be discussed in the following subsection, where response of the source to variations in the gas flow and the coupled microwave power is studied.

In Fig.15, z-dependence of total ion density along the source axis is shown in combination with the ion pressure dependence. The ion pressure is calculated as $P_i(z) = \sum_Q n_{i,Q}(z) T_{i,Q}(z)$, where summation is done over all ion charge states Q. Also, mean ion temperature is shown there as a function of z, being defined as $P_i(z)/\sum_Q n_{i,Q}(z)$. Ion density peaks inside the ECR zone and decreases outside the zone. At the same time, ion temperature gradually increase along z-coordinate outside the ECR zone due to more effective ion heating by the cold electrons with $T_{ec}$=5 eV compared to the heating rate by the warm electrons with $T_{ew}$=2 keV. To the great extent the ion density drop outside the ECR zone is caused by this boost in the ion heating rate. The ion pressure is uniform along the source axis - all irregularities in the ion pressure profile are smoothed by an ion movement along the magnetic field lines.

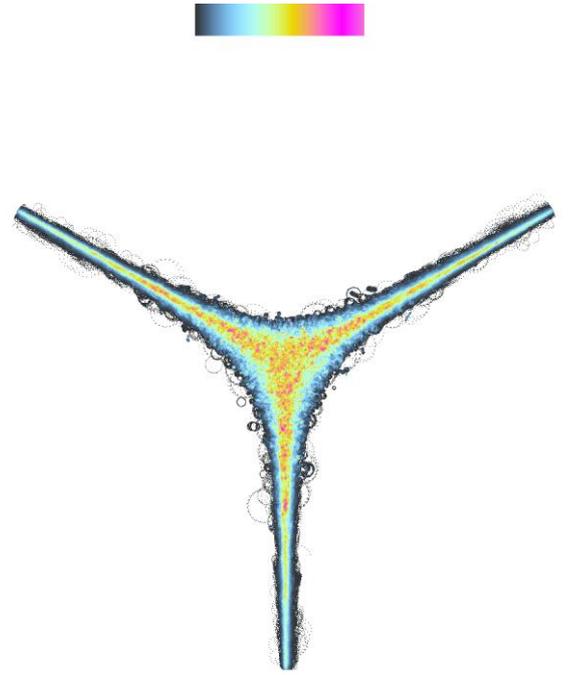

Fig.16. Ion density profile close to the extraction electrode of KVI AECRIS. Trajectories of all charge states are shown

Six-arm spatial distributions of Figs. 9-10 in the plasma center are transformed into three-arm stars close to the terminating flanges at the injection and extraction sides of the source (Fig.16). The reason for the transformation can be understood from symmetries illustrated by Fig.1. These profiles can be directly compared with the sputtering marks at the extraction electrode observed experimentally.

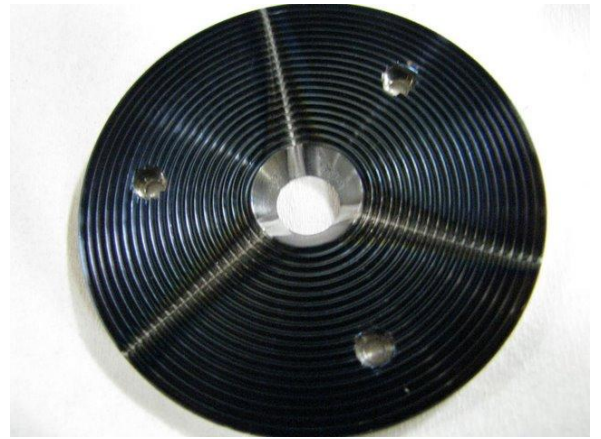

Fig.17. Sputtering marks at the extraction electrode of KVI AECRIS.

The sputtering marks reveal the rather narrow deep trenches along the star arms visible as white lines in the Fig.17. The lines are surrounded with the wider and less pronounced halo. This pattern is clearly seen in the simulated profile. The halo around the narrow strip is due to the lowly charged ions, mostly $Ar^{1+}$ ions.

The ions form the extracted ion beam when leaving the source chamber through the extraction aperture (Ø8-mm for KVI AECRIS). The calculated spatial distributions of the extracted ions for KVI AECRIS are always peaking at the axis. The higher is the charge state of ions, the more they are localized close to the axis. This is illustrated by Fig.18, where the mean radial displacement of the extracted ions in the transversal direction ($\frac{\sum_N \sqrt{(x_i^2 + y_i^2)}}{N}$) is plotted as a function of the ion charge state.

Emittance of the extracted ion beams for ECRIS is mainly determined by the transversal size of the ion spatial distribution. The so-called "magnetic emittance" term of the total emittance value is caused by conservation of the canonical angular momentum of ions during their extraction from the magnetic field of ECRIS. For the uniform initial radial distribution of ions, the normalized emittance can be calculated as

$$\varepsilon_{MAG}^{rms-norm} \approx 0.04 B_0 r_0^2 \frac{Q}{M} \text{ [}\pi\times\text{mm}\times\text{mrad]},$$

where $B_0$ is the magnetic field in Tesla, $r_0$ is the beam initial size in mm and M is the ion atomic mass. It is seen that the beam emittance is directly proportional to the ion charge state. In practice, however, it is sometime measured that the ion beam emittance is decreasing with the ion charge state [29]. The usual explanation for the effect is that ions with higher charge state are localized at the axis of ECRIS such that their initial radial distribution becomes smaller with increasing their charge state Q.

We see from Fig.18 that the extracted beam initial size is indeed decreasing with Q, but not fast enough to make the emittance decreasing with the ion charge state.

For KVI AECRIS, the extracted ion beam profiles had been observed in experiment by using the viewing targets [28]. The triangular shapes had been seen with a rather uniform distribution of beam intensity inside the triangle, in the good agreement with what is illustrated by Fig.16. For other sources, hollow beam profiles are reported [30]. In part, the hollow profiles can be explained by aberrations in optical elements of low-energy beam line during the beam formation and transport. There are indications, however, that the hollow beams are formed inside ECRIS plasma [31].

Hollow beams are seen in our simulations for FLNR sources: $ECR4-M^2$, DECRIS-2M and DECRIS-SC2. For all these sources, a local minimum in the magnetic field radial distribution is observed as shown in Fig.4. An example of such beam profile is shown in Fig.19 for DECRIS-SC2 18 GHz source. The spatial distribution of $Ar^{8+}$ ions is shown there, for a tuning of the source when the $B_{min}$ exceeds the value optimal for the $Ar^{8+}$ ion beam extraction. For the optimal conditions of the source, the profile is not so sharp but remains to be hollow with the ion current density maximum shifted from the axis.

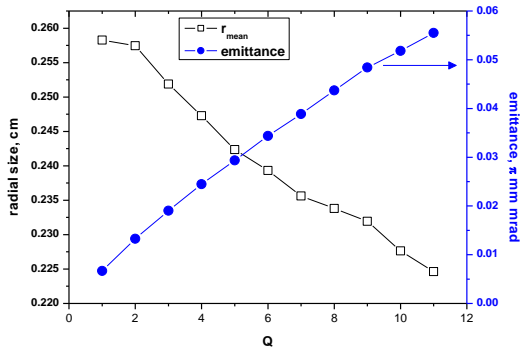

Fig.18. Mean size of the extracted ion spatial distribution in the radial direction and normalized magnetic emittance as a function of ion charge state (right scale). $T_{ew}$=2 keV, 100 W.

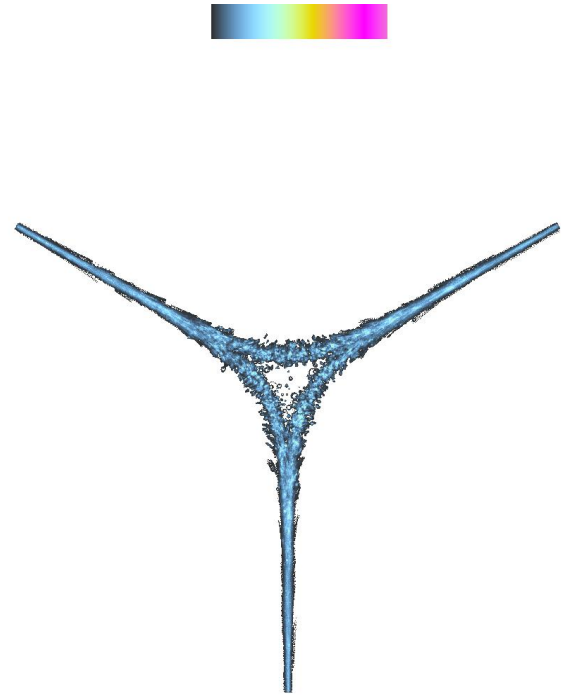

Fig.19. Spatial distribution of $Ar^{8+}$ ions close to the extraction electrode. DECRIS-SC2 18 GHz source. $B_{min}$ value is above the optimum.

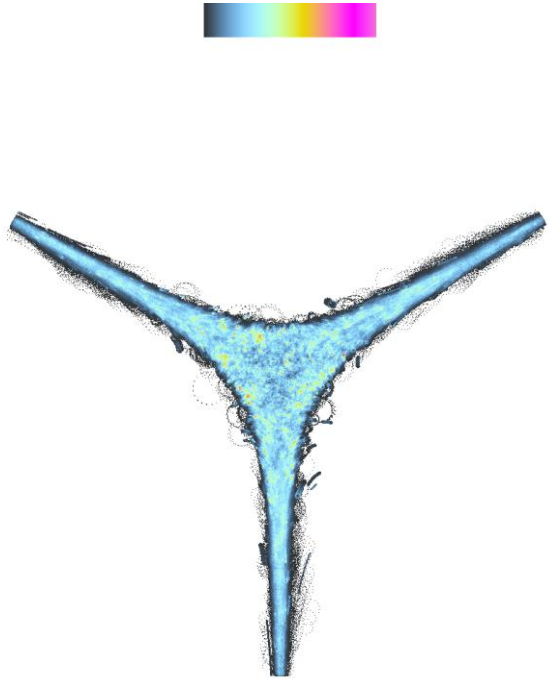

Fig.20. Spatial distribution of $Ar^{1+}$ ions close to the extraction electrode. DECRIS-SC2 18 GHz source.

For the lower charged ions, hole in their spatial distribution is not so pronounced due to the larger Larmor radius of ions and faster diffusion of ions in the radial direction (Fig.20).

Comparing the profiles of Fig.19 and 20, we see that the mean radius of $Ar^{8+}$ extracted ion distribution is larger compared to $Ar^{1+}$ in this specific conditions of the source tuning. However, spread in the distribution of the radial coordinates is smaller for the higher charged ions. To some extent, we have three separate beamlets when extracting such the hollow ion beam and in certain conditions the experimentally measured emittance may be much smaller than the above-mentioned value for the magnetic emittance term.

**C. Dependencies on the gas flow and on the coupled microwave power.**

Dependencies of the extracted ion currents on the gas flow into the system (gas pressure in the plasma chamber) were studied in the following way. First, we select the value of the coupled power. Then, we fix the warm electron temperature ($T_{ew}$) and adjust the statistical weight of computational particles such that in the stationary conditions the coupled power equals to the selected value ±5%. The charge state distribution of the extracted ions is calculated and the process is then repeated by varying the $T_{ew}$ value in the desired range.

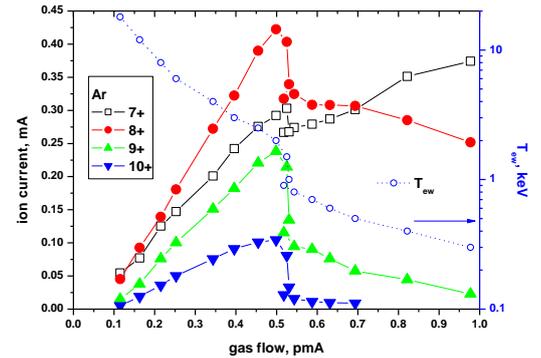

Fig.21. Currents of argon ions as a function of gas flow in/out of the source, left scale. The warm electron temperatures are shown in open blue circles, right logarithmic scale. KVI AECRIS, 100 W of the coupled microwave power. Temperature of the reflected atoms after neutralization of ions at the walls is 0.025 eV.

The total particle flux into the extraction aperture is calculated and is used as a measure of the total gas flow into the plasma chamber. For convenience, we express the gas flow in particle-Amperes, 1 pmA corresponds to the same particle flux as for 1 mA current of singly charged ions. The calculated dependencies are an equivalent of what an operator sees after fixing the injected microwave power at some level and varying the gas flow into the plasma chamber, in assumption that the microwave-plasma coupling coefficient remains the same for the changing plasma conditions.

Variations in the extracted ion currents are shown in Fig.21 for KVI AECRIS and the coupled microwave power of 100 W. The warm electron temperature is also shown in the graph. With increasing gas flow, the electron temperature should decrease to keep the coupled microwave power constant when increasing the particle losses out of the plasma to the source chamber walls and into the extraction aperture. The extracted ion currents grow gradually with the gas flow till the moment when the electron temperature comes below 1.5-2 keV. Below this limiting value, the ion currents drop suddenly; this "high-to-low" (HL) transition is accompanied with increase in the radial size of the highly charged ion distribution (Figs.12 and 14) and with decrease in the electron density at the source axis. Such the transition is experimentally observed when tuning the KVI AECRIS.

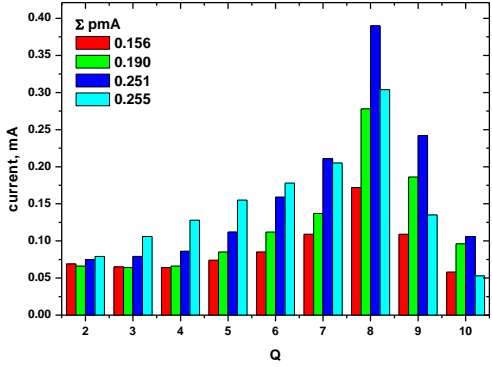

Fig.22. Experimentally measured charge state distributions of the extracted ion currents for different gas flows. KVI AECRIS, 800 W.

The experimentally measured charge state distributions of the extracted argon ion currents are shown in Fig.22 for KVI AECRIS with different gas flows. At this, the injected microwave power was 800 W. The blue bars in Fig.22 represent the maximized output of the source, while cyan bars show the source output when the gas flow into the chamber is slightly above the optimal value. The graphs illustrate the mode jump in the source performance and increase in the source output with increasing the gas flow in the high mode of operation.

Comparing Figs.22 and 21, we see that a ratio between currents in the high and low modes is well reproduced in the simulations. When tuning the source for production of moderately charged ions such as $Ar^{8+}$, most efforts typically is put to increase the gas flow into the chamber without slipping into the low mode; when source is in the low mode, it is necessary to decrease the gas flow substantially to restore the source good performance.

The total particle current out of the source for the charge states (2+-10+) is indicated in the Fig.22. These values are indicators of the gas flow into the plasma chamber to some extent, neglecting the flux due to $Ar^{1+}$ ions and $Ar^0$ atoms. Contribution of these particles is small, neutral atom flux does not exceed (10-15) %. Comparing Figs. 21 and 22, we see that the gas flow at the mode jump is significantly differs from the simulations indicating the over-estimation of the lowly charged ion currents in the model (see also Fig.6).

In Fig.23, mean electron density inside the ECR zone at the source axis is shown as a function of the gas flow into the plasma chamber in the same conditions as in Fig.21. The density is growing steadily with increasing the gas flow, but never exceeds the critical value of $2.5 \times 10^{12}$ cm$^{-3}$ for 14 GHz microwave frequency. Saturation is observed when the electron temperature is close to 3 keV. At the H-L transition, the electron density drops by factor of almost two, and then slowly increasing again till the moment when the electron temperature reaches (300-200) eV.

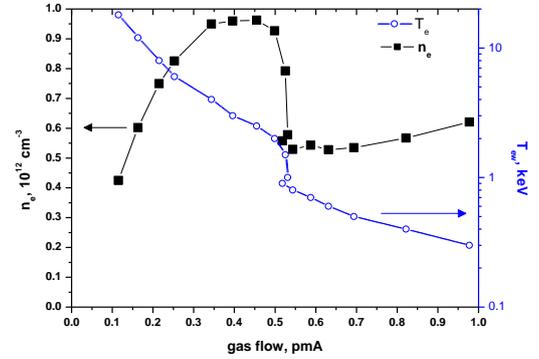

Fig.23. Mean electron density at the source axis inside the ECR zone as a function of gas flow in/out of the source, left scale. The warm electron temperatures are shown in open blue circles, right logarithmic scale. KVI AECRIS, 100 W of the coupled microwave power.

Mean ion temperature in ECR zone is not changing noticeably during the HL transition, remaining at the level of 0.25 eV for $Ar^{8+}$ ions. General trend for the ion temperature is to increase with the increasing gas flow. For the range of gas flow variations depicted in Figs. 21, temperature of $Ar^{8+}$ ions inside the ECR zone increases from 0.1 to 0.4 eV. We refer here to the ion temperature averaged over all ions with the given charge state inside the ECR zone, which gives slightly smaller values compared to the mean ion temperature at the source axis shown in Fig. 15.

This is the electron temperature that defines the conditions for the HL transition. After changing the selected value for the coupled microwave power from 100 W to 200 W, the transition occurs at the higher gas flow but at the same electron temperature. This is illustrated by Fig.24, where the extracted ion current for $Ar^{8+}$ ions is shown as a function of the gas flow for 100 W of the coupled microwave power (open squares) and 200 W (solid red circles). The corresponding electron temperature values are also shown. In the low mode, current of $Ar^{8+}$ ions steadily decreases with the gas flow for 100 W of the coupled power, but increases for the higher power until the electron temperature drops to ≈300 eV, where the local maximum of the current exists. The higher charge state ion currents decreases both for 100 and

200 W in the low mode. In the high mode currents of extracted ions are the same for 100 and 200 W, indicating the source output saturation when increasing the microwave power in the given range.

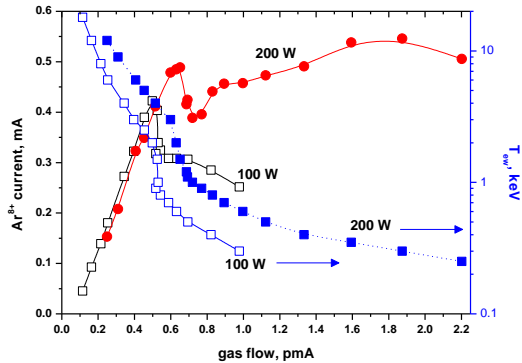

Fig.24. Extracted current of $Ar^{8+}$ ions as a function of the gas flow for 100 W and 200 W of the coupled microwave power. The corresponding electron temperature values are shown (right logarithmic scale). KVI AECRIS.

Sudden change in the extracted ion currents during the HL transition is specific for the cold room temperature gas inside the plasma source chamber. The cold gas condition assumes full accommodation of the excess energy when ions hit the walls and are reflected back into the plasma chamber. If the reflected atoms retain some energy that was acquired in the sheath layer before hitting the wall, then atoms can penetrate deeper into the plasma before being ionized.

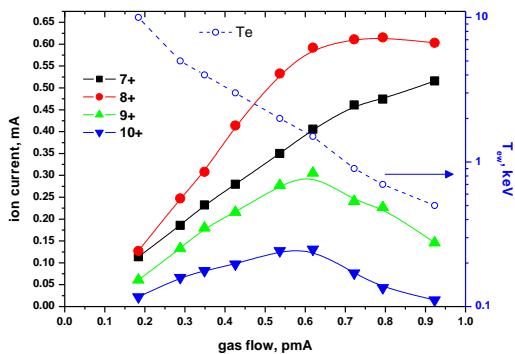

Fig.25. Currents of argon ions as a function of gas flow in/out of the source, left scale. The warm electron temperatures are shown in open blue circles, right logarithmic scale. KVI AECRIS, 100 W of the coupled microwave power. Temperature of the reflected atoms after neutralization of ions at the walls is 0.15 eV.

In Fig.25, the results of calculation of the extracted ion currents are shown as a function of the gas flow into the chamber in assumption that the reflected atoms after ion neutralization have the energy distribution with the temperature of 0.15 eV. It was checked prior the calculations that changes in the temperature of the reflected atom in the range of (0.1-1) eV do not affect the dependencies significantly. The specific value of 0.15 eV was selected on a base of experimental studies of the velocity distributions of neon atoms released out of a carbon surface in contact with hot neon plasma [32] and measurements of argon atoms velocities scattered from a silicon surface under argon keV ion irradiation [33]. Energy of the reflected atoms is supposed to be fully accommodated in their next collision with the walls. For the "warm" atoms, extracted ion currents are significantly higher compared to the case of the "cold" gas; the gain for the $Ar^{8+}$ ions is around 50% for the same gas flow. No high-to-low transition occurs and currents are smoothly increasing with the gas flow up to some level and then decrease with decreasing electron temperature. For $Ar^{8+}$ ions, the optimal electron temperature is around 1 keV in these conditions.

Comparing the dependencies of Fig.21 and 25 with the experimentally observed behavior of KVI AECRIS, we conclude that for this source the "cold" gas model is more appropriate. The wall material for KVI AECRIS is (oxidized) aluminum, and the wall atoms are lighter than the impinging argon ions. Kinematically this presumes the full energy accommodation [24]. The situation can be different for other wall materials, such as a stainless steel used in FLNR sources. In following calculations, however, we assume the full accommodation and cold gas conditions.

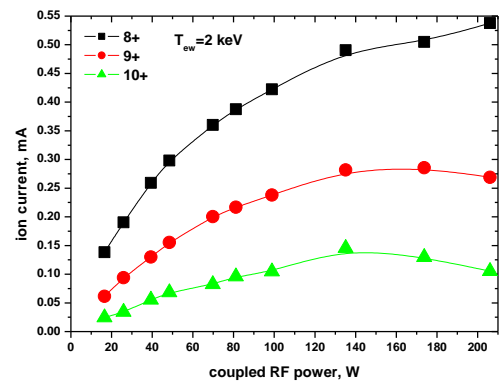

Fig.26. Extracted ion currents at $T_{ew}=2$ keV for different coupled microwave power. KVI AECRIS.

Having the electron temperature fixed at the level of 2 keV optimal for the highly charged ion production, we investigate the dependence of the extracted ion currents on the coupled microwave power. This dependence for $Ar^{8+}$-$Ar^{10+}$ ion currents is shown in Fig.26.

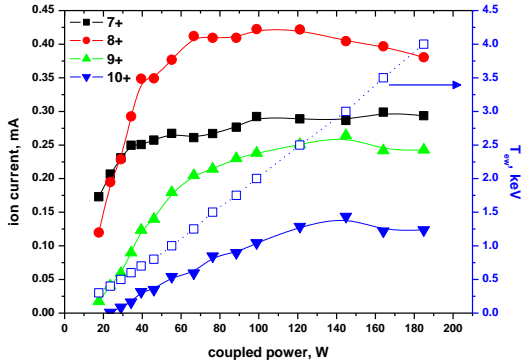

Fig.27. Dependence of the extracted ion currents on the coupled microwave power at the gas flow fixed at (0.5±10%) pmA. Electron temperature is also shown (right scale, open squares).

The extracted ion currents increase with increasing the coupled microwave power up to 150 W and then saturate. Note here that the gas flow should also increase with increasing the microwave power in order to keep the electron temperature fixed at the optimum. From the source operator's point of view, dependence of Fig.26 can be obtained by fixing the injected microwave power and optimizing the gas flow into the plasma source chamber to maximize the highly charged ion currents.

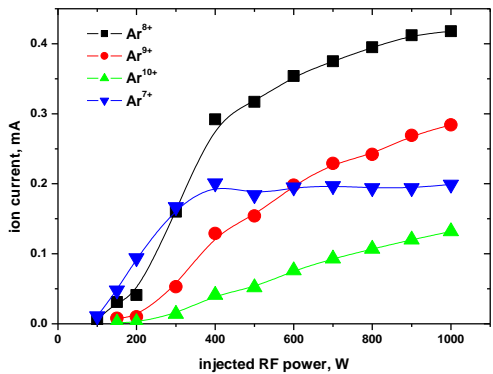

Fig.28. Experimentally measured extracted ion currents for different injected microwave powers. KVI AECRIS, gas flow is fixed at the level optimized to maximize the $Ar^{8+}$ ion current at 800 W of the injected microwave power.

Another way to study the source response to variations in the coupled microwave power is to fix the gas flow and then change the power. The simulated extracted currents for such situation are presented in Fig.27. The gas flow is selected to be 0.5 pmA, which corresponds to the source parameters tuned to maximize the $Ar^{8+}$ current at 100 W of the coupled power. With the fixed gas flow, changes in the coupled power are due to changes in the electron temperature, which vary from 500 eV to 4 keV for the depicted range of the power variation.

The simulated dependencies can be directly compared with the experimental data of the ion current response to changes in the injected microwave power (Fig.28) for KVI AECRIS. The general trend is well reproduced in the simulations, with fast increase in the currents followed by the saturation of low charged ion currents at above 500 W for the experimental dependence and 50 W for the simulations. From the dependencies, we may conclude that the injected microwave power of 1 kW roughly corresponds to 100 W of the coupled power for KVI AECRIS. This means that the microwave power coupling coefficient is around 10% for the given source. The value looks small but it is worth to remind here that the coupled power is calculated without taking into account the non-Maxwellian shape of the electron energy distribution function. Energetic tails of EEDF can carry away as much power as the main body of the distribution - the coupled power is defined in the simulations with a possible error of two or so.

All in all, the simulations agree well with the experimentally observed behavior of the source when varying the gas flow into the plasma and the injected microwave power.

### D. Ion confinement

The longer ions stay in the plasma with a given electron temperature and density, the higher is their chance to be ionized into the higher charge states. Residence (confinement) times define a shape and intensity of the charge state distributions of the extracted ion currents. To measure these times, we calculate the time interval for each individual ion between the moment of its creation during ionization of argon atom and the moment when the ion hits the extraction aperture being in some charge state [7].

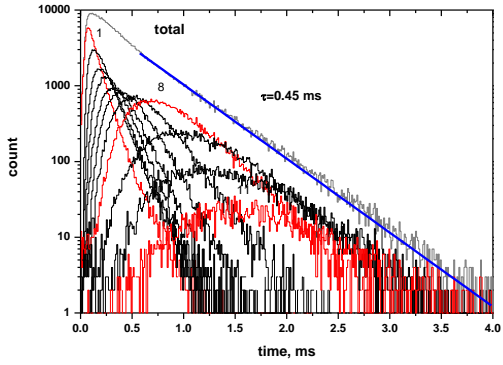

Fig.29. First passage time distributions for the extracted ions. KVI AECRIS, $T_{ew}=2$ keV

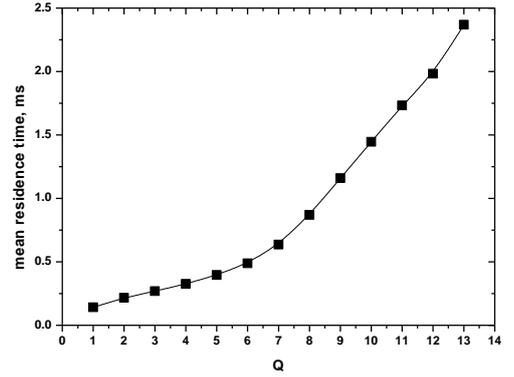

Fig.30. Mean values of FPTD as a function of the ion charge state.

Distributions of these times are the First Passage Time Distributions (FPTD) widely used in theories of a random walk movement. Typical distribution for KVI AECRIS is shown in Fig.29 for the conditions presented in Fig.6 and optimized to produce $Ar^{8+}$ ions. The total distribution is shown that is constructed summing the passage times of ions in all charge states, combined with the charge state resolved distributions. The resolved distributions are shown for argon ions from (1+) to (11+), with the maximum of distributions shifting to the larger times with increasing the charge state. Some distributions are labeled in the graph with their charge state and are colored red. The total distribution's shape is well described at the times ≥0.1 ms by the exponential decay curve. Time constant of the curve is 0.45 ms for the given conditions.

Mean values for the charge-resolved FPTDs are shown in Fig.30. Linear increase in the mean passage times is seen for the highest charge states. These times can be understood as the breeding times. It is requested around 2 ms for argon ions to reach the charge state of 13+ in the given conditions of the plasma. Only a small fraction of all ions reach this charge state; this fraction depends on how fast the ions are leaving the plasma.

The ion confinement time is defined by the decay constant of the total FPTD, the time does not depend on the ion charge state at least for the highest states.

Experimentally, the ion confinement times in ECRIS plasma were estimated for argon ions with Q≥9 by comparing the ion densities $n_q$ in plasma and the extracted ion currents $I_q$ [5]. These times were calculated by Douysett et al. [5] as

$$\tau_q = \kappa \frac{(2L)S}{2} \frac{n_q q e}{I_q}$$

where 2L is the estimated plasma length, L is the ECR zone length, S is the extraction aperture area and κ is a transmission efficiency of the beam line. Uncertainty in the confinement times was quoted to be within a factor of two.

We make the same estimations by using the ion densities along the source axis inside the ECR zone as presented in Fig.15. In our case, the ECR zone length is L=10.8 cm and S=0.5 cm2 for the Ø8-mm extraction aperture. The confinement times calculated as quoted above are presented in Fig.31. Saturation is observed for Q higher than 8+ at the level of 0.5 ms, which is close to the FPTD decay time constant. For the lower charge states, the times grow linearly with Q indicating the ion "losses" due to ionization into the higher charge states.

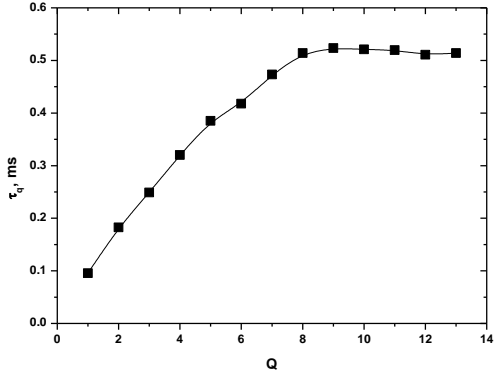

Fig.31. Confinement times of argon ions.

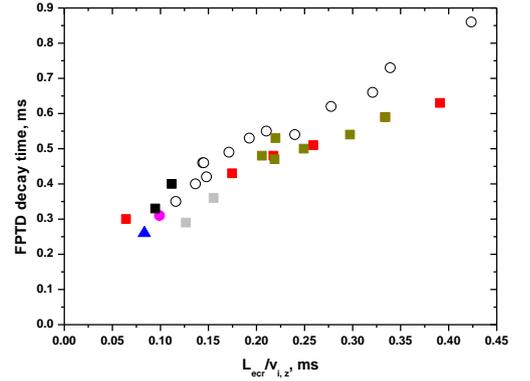

Fig.32. Correlation between the FPTD decay time constant and the scaling factor $\frac{L_{ECR}}{\langle|v_{iz}|\rangle}$.

The decay constants of the FPTD depend on the ion temperature and the ECR zone length. We calculated these constants for all sources that were simulated by the model with different gas flows and coupled powers. The KVI AECRIS data are presented for three configurations of the source – default, $B_{min}$ and flat configurations. Parameters of two last configurations will be discussed in the next subsection. Combined all together, the constants are presented in Fig.32 as a function of the scaling factor that is calculated as $\frac{L_{ECR}}{\langle|v_{iz}|\rangle}$ , where $L_{ECR}$ is the ECR zone length along the source axis and $\langle|v_{iz}|\rangle$ is a mean magnitude of $Ar^{8+}$ ion velocities inside the ECR zone along z-axis. The $Ar^{8+}$ ions are chosen because they are the main contributors into the FPTD for the times where the fit to the FPTD shape is done (≥0.5 ms). Anyway, the ion temperatures do not change much for the highly charged ions due to their effective temperature equilibration in ion-ion collisions. Only one spatial component of ion velocities is taken into account and it was checked out that as expected for the isotropic velocity vectors $\langle|v_{iz}|\rangle = \frac{\langle|v_i|\rangle}{\sqrt{3}}$ , where $\langle|v_i|\rangle$ is the mean ion velocity.

Linear correlation between the decay constant and scaling factor is observed:

$$\tau_D \cong 2\frac{L_{ECR}}{\langle|v_{iz}|\rangle}$$

What is seen in Fig.32 is an indication of the so-called gas-dynamical ion confinement known from a theory of open-ended mirror traps. There, the ion confinement times are estimated as ~ $\frac{RL}{v_{Ti}}$ , where R is the magnetic field mirror ratio R=$B_{max}/B_0$ in a trap [34]. This estimation of the ion confinement times assumes that the plasma length is much larger than the ion-ion collision length, which is a true for the ECRIS plasmas.

We conclude therefore that the gas-dynamical confinement of ions is sufficient to explain the experimentally observed extracted ion currents out of ECRIS.

## IV. Magnetic field scaling

Magnetic configuration greatly influences an ECRIS performance. When designing the sources, the empirical scaling [35] is used to select the characteristic values of the magnetic field profile

$$\frac{B_{INJ}}{B_{ECR}} \geq 4, \quad \frac{B_{EXT}}{B_{ECR}} \geq 2$$

$$\frac{B_H}{B_{ECR}} \geq 2, \quad \frac{B_{Min}}{B_{ECR}} \sim 0.8$$

Here, $B_{inj}$ and $B_{ext}$ are the magnetic field maxima at injection and extraction sides of a source at the source axis, $B_{min}$ is minimal magnetic field at the axis and $B_H$ is the hexapole field at the source chamber radial wall. The values are normalized to $B_{ecr}$, the field that corresponds to the electron cyclotron resonance. This scaling was obtained experimentally by varying the magnetic field configurations and optimizing the highly charged ion output. Output currents are either strongly peaked or saturated at the optimal values of the magnetic field.

In this section, we report on changes in the extracted ion currents as simulated by our model for different profiles of the magnetic field. KVI AECRIS is modeled in most details, with an emphasis put on the $Ar^{8+}$ currents. Behavior of highly charged ions with $Q \geq 9$ closely resembles the $Ar^{8+}$ tendencies. Results for other sources listed in Table I will be given to compare with KVI AECRIS when necessary. Data are obtained with fixing the electron temperature at 2 keV for KVI AECRIS (3 keV for other sources).

We begin with studying the $B_{min}$ dependencies of the extracted currents. The newest ECR ion sources normally use a middle coil to control the $B_{min}$ value of the field distribution without affecting the $B_{inj}$ and $B_{ext}$ values significantly. KVI AECRIS design does not have such the coil (Fig.2). We calculate the magnetic field distribution for KVI AECRIS by inserting a middle coil in between the injection and extraction coils. When energizing the coil, $B_{min}$ value can be controlled with changing the magnetic field at the extraction and injection sides of the source by less than 5%.

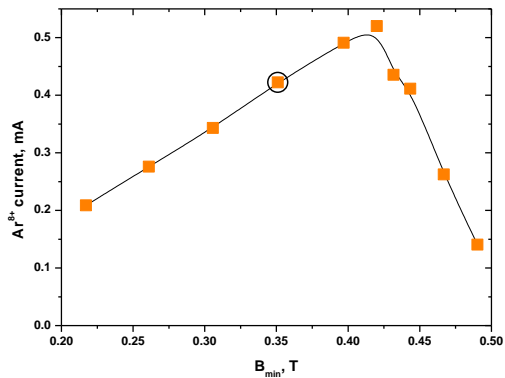

Fig.33. Extracted current of $Ar^{8+}$ ions for different values of B-min at the source axis. KVI AECRIS, $T_{ew}$=2 keV, 100 W.

Dependence of $Ar^{8+}$ ion currents on $B_{min}$ is shown in Fig.33. Currents in the injection and extraction coils are 1 kA corresponding to the maximal achievable values of $B_{inj}$ and $B_{ext}$. The highlighted point at the graph corresponds to the default $B_{min}$ value without energizing the middle coil - an equivalent of not having the coil in the magnetic design.

Calculations are done with setting the coupled power to 100 W. To calculate each data point depicted in Fig.33, the particle statistical weight is varied such as to keep the power at the level of 100 W, which implies that for the fixed electron temperature the total flux of ions/electrons out of the plasma is kept constant. Flux of the particles through the extraction aperture is changing when varying the $B_{min}$. From point of view of experimentation, data points in the Fig.33 can be obtained by optimizing the gas flow each time when changing the $B_{min}$ value (the optimal gas flow increases with the increasing $B_{min}$).

Optimal value of $B_{min}$ is close to 0.8 in agreement with the scaling laws.

The main effect of the varying $B_{min}$ is a strong variation in the ECR zone size both in radial and axial directions. When $B_{min}$ equals to $B_{ECR}$, the ECR zone is reduced to a point and ECR-heated plasma cannot be produced. For the range of $B_{min}$ variations in Fig.33, the ECR zone length ($L_{ECR}$) changes from 14 (for the low $B_{min}$) to 3 cm. Reduction of $L_{ECR}$ affects the ion confinement time in the way as seen in Fig.32. Distinctively, larger radial size of the ECR zone blocks penetration of the neutral particles toward the source axis through the dense parts of the plasma, which decreases the extracted ion current. Also, smaller magnetic field boosts the plasma diffusion across the magnetic field lines reducing the plasma density and increasing the plasma radial size. Counteraction of these tendencies results in existence of an optimal value for $B_{min}$.

These considerations were checked by performing two separate calculations: first, we switched off the plasma diffusion process due to electron-ion collisions by making the ion heating isotropic. The result is the denser plasma on the source axis and higher extracted ion currents. Current of $Ar^{8+}$ ions reached the level of 1.1 mA for the optimal $B_{min}$ of around 0.32 T; decrease of the extracted current with smaller $B_{min}$ is not so pronounced now and current of $Ar^{8+}$ is around 0.8 mA for the smallest investigated $B_{min}$ shown in Fig.33. This should be compared with the drop by factor of 2.5 in the $Ar^{8+}$ current in the normal conditions. Above the optimal $B_{min}$, the currents still decrease for the simulations with the plasma diffusion switched off.

The counteracting effect was studied by changing the magnetic field such that the magnetic field gradient in the axial direction is small inside the ECR zone. This is so-called flat magnetic field profile. Experimentally, this was demonstrated [36] to be an effective method to increase the extracted ion currents. In our simulations, to obtain such the profile it was necessary to calculate the magnetic field with installing two axially symmetric soft-iron rings in addition to the middle coil and two default solenoids.

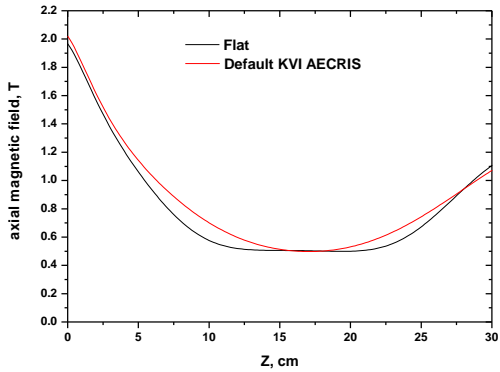

Fig.34. Axial magnetic field for the flat and normal configurations of KVI AECRIS.

In Fig.34, the calculated profile of the axial magnetic field is shown in combination with the default profile for KVI AECRIS. With the flat profile, size of the ECR zone in axial direction for the same $B_{min}$ is much larger compared to the normal profile. In the radial direction, the ECR zone size for the given $B_{min}$ is approximately the same for the flat and normal profiles being mostly determined by the hexapole field.

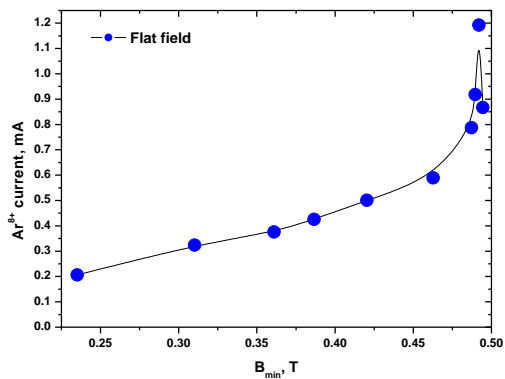

Fig.35. Extracted current of $Ar^{8+}$ ions for different $B_{min}$ at the source axis. KVI AECRIS, flat profile of the magnetic field, $T_{ew}$=2 keV, 100 W.

Dependence of $Ar^{8+}$ ion current on the $B_{min}$ value for the flat magnetic field is shown in Fig.35. The same procedure was used as for getting the data in Fig.33 – the electron temperature was set to 2 keV, the coupled power was 100 W throughout the calculations and the particle statistical weight was varied to keep the values constant.

The optimal value of $B_{min}$ is very close to the resonance for the flat profile of magnetic field. Current of $Ar^{8+}$ ions reaches 1.2 mA, much higher than for the normal profile. Current decreases for $B_{min}$ above 0.49 T; difference is small between the flat and normal profile values for $B_{min}$ below 0.4 T.

Comparing the results shown in Figs.33 and 35, we conclude that the flat profiles of the axial magnetic field are preferable when the $B_{min}$ values are close to the resonance and that the currents decrease for the high $B_{min}$ in the normal configuration because of a reduction of the ECR zone axial size.

For KVI AECRIS, it is not possible to compare the simulated currents for different $B_{min}$ values with the experiment. It is possible to do for DECRIS-SC2 source, which is equipped with the middle coil used to control the $B_{min}$ values. The middle coil is energized in the direction opposite to others and the increasing current in the coil decreases the $B_{min}$ value.

Current of $Ar^{8+}$ ions was measured for DECRIS-SC2 18 GHz source as a function of a current in the middle coil. The injected microwave power was fixed at 300 W, and current in the injection/extraction coils was set to 70 A close to the maximum value of 75 A. Current of $Ar^{8+}$ ions is shown in Fig.36 as solid black squares. Gas flow was not varied when measuring the extracted currents.

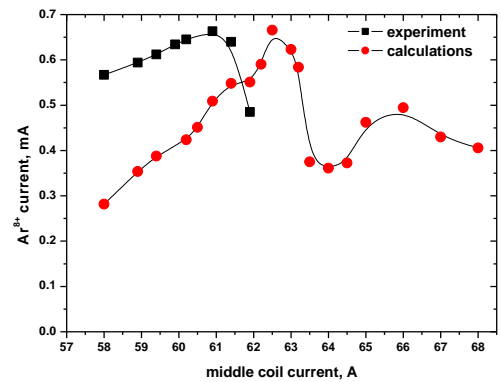

Fig.36. Experimental and simulated dependencies of $Ar^{8+}$ ion current on a current in the middle coil of DECRIS-SC2 18 GHz source.

In the Fig.36, the simulated currents of $Ar^{8+}$ ions are shown as red circles. The simulations are done with fixing the coupled microwave power at the level of 200 W and for the fixed gas flow out of the source equal to (0.8±5%) pmA. The electron temperature $T_{ew}$ reached 3 keV at the maximum of the shown dependence (at 62.5 A of the middle coil current) and decreased in both directions from the maximum – it reached 2.3 keV for the minimal investigated current in the coil and 1.1 keV at the maximal current in the coil. The $B_{min}$ value is maximal at the lowest current in the middle coil ($B_{min}/B_{ECR}$=0.81, $B_{ECR}$=0.643 T) and minimal at the highest current ($B_{min}/B_{ECR}$=0.64). Current of $Ar^{8+}$ ions is maximized at $B_{min}/B_{ECR}$=0.73.

For current in the middle coil above the optimum, high-to-low mode transition occurs resulting in a fast drop in the extracted ion currents. This jump in the currents has essentially the same origin as the jump seen in Fig.21, with a decrease in the electron temperature and an increase in the plasma radial size inside the ECR zone. Other branch in the dependence of Fig.36 for the lower coil current is characterized by an increased hollowness of the extracted ion beam. The beam shape typical for these coil currents is shown in Fig.19; for the higher axial magnetic field, the radial local minimum in the field starts to be more pronounced and the plasma is more localized there.

Comparing the maxima in the experimentally measured and simulated dependencies, we see that the simulations differ from the experiment by 2 A in the coil current, which is less than 5%. Agreement is sufficiently good also in the absolute values of the extracted ion currents.

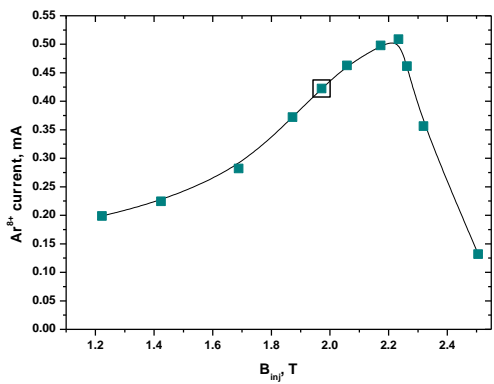

Fig.37. Currents of $Ar^{8+}$ ions for different magnetic fields at the injection side of the source. KVI AECRIS, $T_{ew}$=2 keV, 100 W.

Dependencies of extracted ion currents are also obtained for varying separately the magnetic fields at the injection and extraction sides of KVI AECRIS. The fields are changed by changing the currents in the injection and extraction coils of the source (Fig.2 and 3). When changing the currents, not only the field at the corresponding side of the source is changing, but also $B_{min}$ and the axial and radial ECR zone sizes. Fields at the opposite sides are not affected by variations in the coil currents significantly.

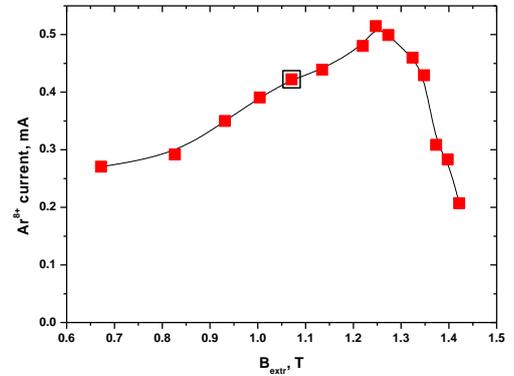

Fig.38. Currents of $Ar^{8+}$ ions for different magnetic fields at the extraction side of the source. KVI AECRIS, $T_{ew}$=2 keV, 100 W.

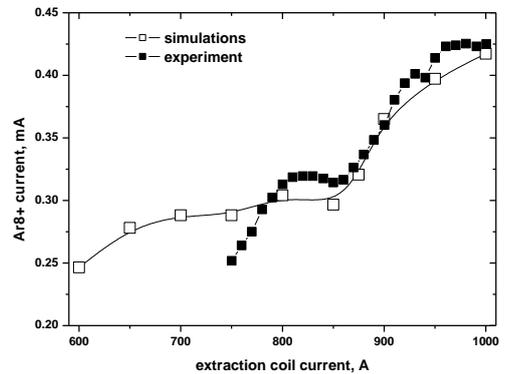

Fig.39. Simulated and experimental dependencies of $Ar^{8+}$ ion current on the extraction coil current. Gas flow is constant (0.5 pmA), current in the injection coil is 1 kA. KVI AECRIS, 100 W.

Changes in the extracted current of $Ar^{8+}$ ions are shown in Fig. 37 for the injection magnetic field variations and in Fig.38 for the extraction magnetic field variations. The dependencies are obtained with a constant electron temperature of 2 keV and for the coupled power of 100 W, in the same way as for Figs. 33 and 35. The data points for the default currents (1 kA) in the extraction/injection coils are highlighted in the graphs.

For the variation in the injection coil current depicted in Fig.38, $B_{min}$ changed from 0.25 to 0.48 T. Also, the

ECR zone length at the source axis decreases from 15.7 cm to 3 cm. For the dependence in Fig.39, the variation in $B_{min}$ is from 0.26 to 0.47 T, and the ECR zone length changes from 16.2 to 4.5 cm for the depicted range of the extraction coil current variation. We see that it is a rather difficult to disentangle the effects connected to the changes in the injection/extraction fields and to the changes in $B_{min}$ value as illustrated in Fig.33. The dependencies look the same: extracted ion current of $Ar^{8+}$ grows with increasing the injection/extraction fields and then drops above certain level of the field. The reasons for such behavior are the same as discussed above for $B_{min}$ case – the lost confinement of ions for too short ECR zone and the increased radial diffusion of the particles when mean magnetic field inside the ECR zone is too low. The optimal magnetic fields are 2.2 T ($\frac{B_{INJ}}{B_{ECR}} = 4.4$) at the injection and 1.2 T at extraction ($\frac{B_{EXT}}{B_{ECR}} = 2.4$), close to what is predicted by the empirical scaling laws.

At the lowest investigated current in the injection coil (Fig.38), the injection and extraction magnetic fields are comparable and the field profile is almost symmetric. It is instructive to compare the ion fluxes coming to the injection and extraction sides of the source in this situation. For the injection field of 1.2 T, current of $Ar^{8+}$ ions into the (virtual) aperture of 8-mm diameter at the injection side is 0.17 mA compared to 0.2 mA at the extraction side. For the default configuration of the magnetic field, when the magnetic field at the injection is two times higher than at the extraction, current into the extraction aperture is larger by factor of two compared to the current at injection side – 0.4 mA vs. 0.2 mA for $Ar^{8+}$ ions.

Dependence of $Ar^{8+}$ current on the extraction field is also calculated with the constant gas flow into the source. Again, the electron temperature should be varied to keep the gas flow and the coupled power at the constant level. The calculations are compared to the experimentally measured currents for KVI AECRIS in Fig.39. In the experiment, $Ar^{8+}$ ion current drops when decreasing the extraction coil current and extraction field. The decrease in the ion current is slowed down at the coil current around 800 A and then the ion current continues to decrease with the decreasing coil current. The calculated dependence reproduces the experiment closely. It is obtained by fixing the gas flow at the level of 0.5 pmA and for the constant coupled power of 100 W. The electron temperature is decreasing in the simulations from 1.7 keV at the maximal current in the extraction coil of 1 kA to 500 eV at the lowest coil current. Because of the temperature drop, the high-to-low transition in the plasma occurs when decreasing the extraction coil current below 900 A. Agreement between the calculations and experiment is good.

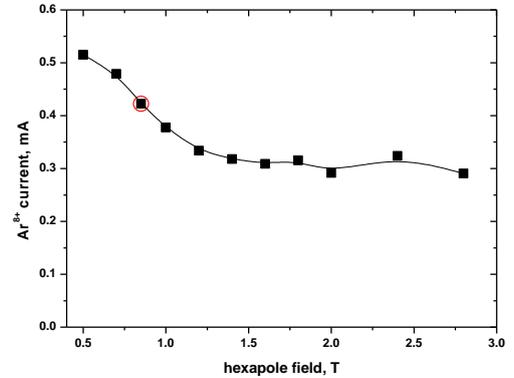

Fig.40. Simulated dependence of $Ar^{8+}$ ion current on the hexapole field. Electron temperature is 2 keV, currents in the injection and extraction coils are 1 kA. KVI AECRIS, 100 W.

Extracted current of $Ar^{8+}$ is calculated for different hexapole fields (Fig.40). Injection and extraction coils are fully energized with 1kA current and the electron temperature is kept constant at the level of 2 keV. The coupled power is kept constant and equal to 100 W. The data point for the default 0.85 T of the hexapole field is highlighted in the graph. For KVI AECRIS ion currents increase with decreasing the hexapole field down to 0.5 T level. For the hexapole field higher than the default one, ion current first decreases and then saturates for the fields higher than 1.5 T.

Higher hexapole field compresses the ECR zone in the radial direction without changing its axial size. Simultaneously, close to the extraction flange the plasma is compressed in three directions in between the star arms and is decompressed along the arm directions (see Fig.16). If the plasma size in between the star arms is much larger than the extraction aperture, stronger hexapole can increase the extracted ion currents. If plasma size and extraction aperture fit each other, any increase in the hexapole strength results in decreasing ion current out of the source. We see two tendencies counteracting each other – smaller radial size of the plasma helps to increase the extracted ion currents, because it is easier for atoms to reach central plasma regions, stronger hexapole redirects more ions from the plasma axis toward the plasma chamber walls.

Increase in the extracted ion current with decreasing the hexapole field below 1 T level is specific for KVI AECRIS. We also perform the calculations for other sources. The results are presented in Fig.41 for ECR4-M$^2$, DECRIS-2M and DECRIS-SC2 14 (18) GHz. For all sources, we fix the coupled power to 200 W and the electron temperature is chosen to be 3 keV optimal for output of the FLNR sources. Axial magnetic field calculations are done for the currents in the extraction and injection coils that maximize the Ar$^{8+}$ ion current as observed in measurements. The corresponding magnetic field values are given in Table 1. The data points are highlighted that correspond to the default hexapole fields of the sources.

For these sources, low hexapole field results in a decrease in Ar$^{8+}$ ion currents, saturation in the currents for the high hexapole fields for DECRIS-SC2 (14-18 GHz) is the same as for KVI AECRIS. Saturation for ECR-4M$^2$ and DECRIS-2M sources is observed for the larger hexapole fields compared to other sources. Decrease in the currents at low hexapole strength is most pronounced for DECRIS-SC2 18 GHz source, which is tuned to the higher axial magnetic fields compared to DECRIS-SC2 14 GHz. Beams of the extracted ions become to be hollow for the low hexapole fields for all FLNR sources.

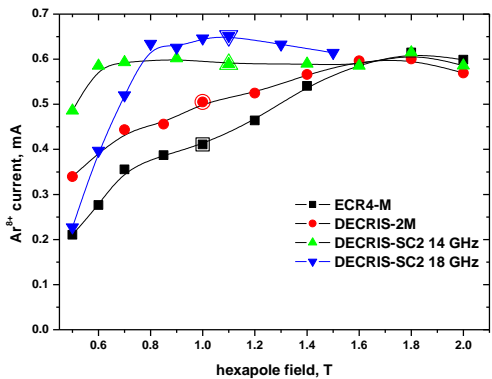

Fig.41 Simulated dependencies of Ar$^{8+}$ ion current on the hexapole field for FLNR sources. Electron temperature is 3 keV, coupled power is 200 W for all sources.

As it was discussed earlier, KVI AECRIS has the lower radial gradient of the solenoidal magnetic field compared to the FLNR sources (Fig.4). When the gradient is relatively large and hexapole field is lower of the default value, the local minimum of magnetic field away from the source axis becomes be more pronounced. For the low hexapole fields, maximal density of the plasma is not on the axis and hollow ion beam is formed, which results in a drop of the extracted ion currents.

Both saturation and decrease of the extracted ion currents with increasing the hexapole field above certain level are observed in a practice [4,37]. Specific response of the sources depends on the magnetic field profile in radial direction according to our calculations. Sources with the larger diameter of the plasma chamber (larger bore of the hexapole magnets) needs in stronger hexapole to suppress the plasma radial instability, as seen from comparison of ECR-4M (Ø7.4 cm) and DECRIS-2M (Ø6.4 cm) sources.

Scaling laws for ECRIS are well reproduced with our code.

## V. Discussion and conclusions

The ECRIS plasma is modeled as a dense hot ($T_{ew}$≈1 keV) plasmoid located inside the ECR zone and separated from the vacuum chamber walls by tenuous cold ($T_{ec}$≈5 eV) plasma. Terms "hot" and "cold" refer here to the electron temperature; contrary, the ion temperature is smallest inside the ECR zone and is maximized in the outermost parts due to the high collisionality of the cold electrons. The halo plasma flattens the ion pressure spatial distribution and slows down the plasma diffusion along the magnetic field lines. Even with no retarding electric fields, lifetime of ions inside the plasma is long enough to explain the experimentally observed extracted ion currents.

Charge-change collisions of the ions with atoms are not important for the ion balance in the plasma due to a relatively small concentration of atoms. Recombination processes are slow and play no role for the argon ions with the charge states below (16+).

The plasma suffers from the gas starvation and any reasonable means to facilitate the atom penetration into the dense parts of the plasma are beneficial for the ion production. Optimized magnetic field profile minimizes the ECR zone radial size while keeping the zone long along the source axis. Flat magnetic field profile looks the best choice.

The optimized plasma has maximal density along the source axis. Depending on the radial gradients of the magnetic field, hollow plasma density profiles can be obtained resulting in the hollow ion beams extracted from ECRIS. Hexapole field, plasma chamber diameter and the solenoidal magnetic field profile should be chosen such as to minimize a possibility for such beam formation; stronger hexapole field does not automatically provide the best source performance because this component of the magnetic field redirects particles away from the extraction aperture of the source.

In the model we assume that electron and ion losses out of the plasma are always equal each other. A link to the processes that equilibrate these losses is missed. In simple bounded plasmas, positive plasma potential builds up to regulate the electron fluxes to the walls. At this, electrons are retarded by electric field inside thin sheath layer close to the walls. The mechanism works for the cold electrons in the halo plasma, but cannot affect directly the ECR-zone mirror-trapped warm electrons. Warm electron losses are determined by at least two processes: the classical scattering into loss-cone due to the electron-ion collisions [3] and burst-like losses due to the electron micro-instabilities [38,39]. The second term in the electron losses is difficult to calculate; it depends on details of the electron energy distribution across and along the magnetic field, plasma density, cavity properties of the plasma chamber and other parameters. We leave the problem of the ion and electron losses equilibration open for further investigations, claiming that ECRIS performance is well described in a guess that no sizeable electric field affects the ion dynamics inside the plasma.

## Acknowledgements

One of us (VM) would like to express his special thanks to his former colleagues at Kernfysisch Versneller Instituut, University of Groningen, for their long-term support. The research was partially supported by RFBR Grant No. 13-02-12011 ofi_m.